\documentclass{article} 
\usepackage[table]{xcolor}
\usepackage{iclr2024_conference,times}

\usepackage{amsmath,amsfonts,bm}

\def\eqref#1{equation~\ref{#1}}

\def\1{\bm{1}}

\DeclareMathAlphabet{\mathsfit}{\encodingdefault}{\sfdefault}{m}{sl}
\SetMathAlphabet{\mathsfit}{bold}{\encodingdefault}{\sfdefault}{bx}{n}

\usepackage{microtype}
\usepackage{hyperref}
\usepackage{url}
\usepackage{booktabs}
\usepackage{graphicx}
\usepackage{natbib}
\usepackage{wrapfig}
\usepackage{multirow}
\usepackage{pifont}
\usepackage{tabularray}
\usepackage{tabularx}
\usepackage{subcaption}
\usepackage{enumitem}

\title{WavCraft: Audio Editing and Generation with Large Language Models}

\author{
    \textbf{Jinhua Liang\textsuperscript{1}},~
    \textbf{Huan Zhang\textsuperscript{1}},~
    \textbf{Haohe Liu\textsuperscript{2}},~
    \textbf{Yin Cao\textsuperscript{3}},~
    \textbf{Qiuqiang Kong\textsuperscript{4}},~
    \textbf{Xubo Liu\textsuperscript{2}},~ \\
    \textbf{Wenwu Wang\textsuperscript{2}},~ 
    \textbf{Mark D.~Plumbley\textsuperscript{2}},~
    \textbf{Huy Phan\textsuperscript{5}\thanks{The work does not relate to H.P.’s position at Amazon.}},~
    \textbf{Emmanouil Benetos\textsuperscript{1,6}}~ \\
    $\textsuperscript{1}$ Centre for Digital Music (C4DM), Queen Mary University of London, UK \\ 
    $\textsuperscript{2}$ Centre for Vision, Speech and Signal Processing (CVSSP), University of Surrey, UK \\
    $\textsuperscript{3}$ Xi’an Jiaotong Liverpool University \hspace{0.2cm}
    $\textsuperscript{4}$ The Chinese University of Hong Kong \\
    $\textsuperscript{5}$ Amazon, Cambridge, MA, USA \hspace{0.75cm} 
    $\textsuperscript{6}$ The Alan Turing Institute, UK \\
    \texttt{jinhua.liang@qmul.ac.uk}
}

\iclrfinalcopy 
\begin{document}

\maketitle

\begin{abstract}
We introduce WavCraft, a collective system that leverages large language models (LLMs) to connect diverse task-specific models for audio content creation and editing. Specifically, WavCraft describes the content of raw audio materials in natural language and prompts the LLM
conditioned on audio descriptions and user requests. WavCraft leverages the in-context learning ability of the LLM to decomposes users' instructions into several tasks and tackle each task collaboratively with the particular module. Through task decomposition along with a set of task-specific models, WavCraft follows the input instruction to create or edit audio content with more details and rationales, facilitating user control. In addition, WavCraft is able to cooperate with users via dialogue interaction and even produce the audio content without explicit user commands. Experiments demonstrate that WavCraft yields a better performance than existing methods, especially when adjusting the local regions of audio clips. Moreover, WavCraft can follow complex instructions to edit and create audio content on the top of input recordings, facilitating audio producers in a broader range of applications. Our implementation and demos are available at this \href{https://github.com/JinhuaLiang/WavCraft}{https://github.com/JinhuaLiang/WavCraft}.
\end{abstract}

\begin{figure*}[h]
    \centering
    \includegraphics[width=1.0\linewidth]{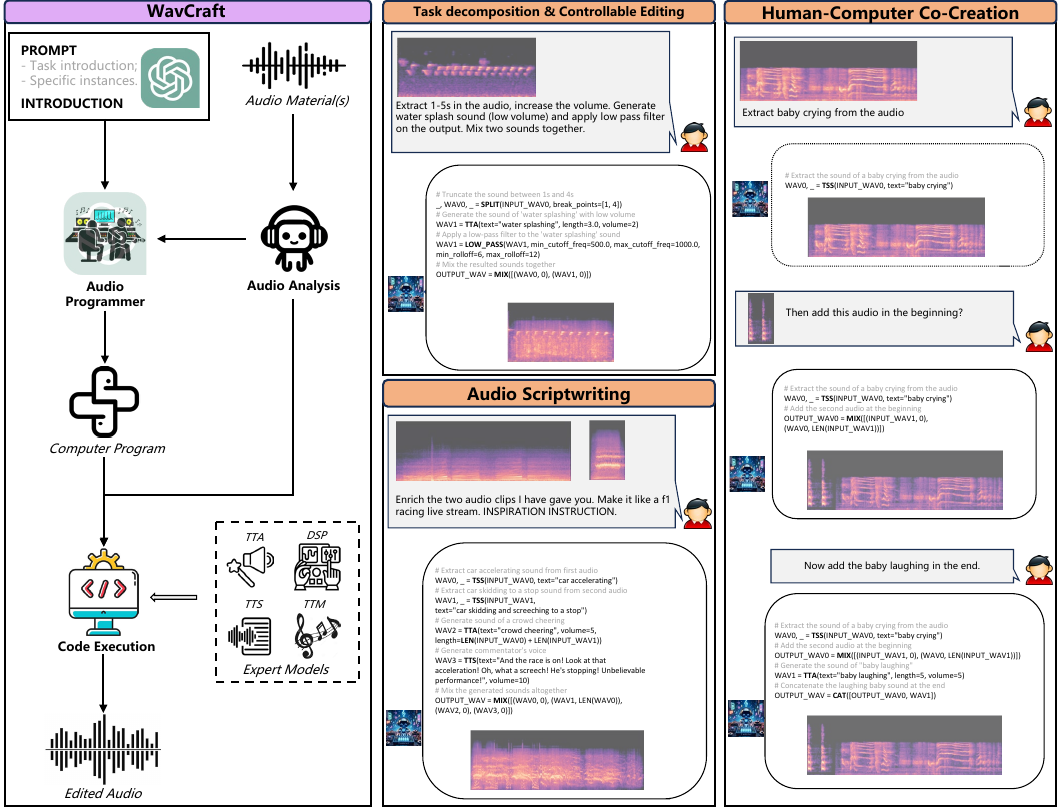}
    \caption{WavCraft overview. WavCraft processes the content of input audio clips and prompts the LLM to generate the code conditioned on user query and audio content. The generated code is then implemented as a computer program empowered by a set of expert models. WacCraft is capable to tackle cases involving: 1) task decomposition; 2) controllable editing; 3) human-computer co-creation; and 4) audio scriptwriting.}
    \label{fig:wavcraft_overview}
\end{figure*}

\section{Introduction} \label{sec:introducation}
Large language models (LLMs) such as ChatGPT~\citep{openai_gpt-4_2023} have remarkably promoted the development of artificial intelligence-generated content (AIGC). Driven by large-scale pre-training on massive high-quality textual tokens and reinforcement learning from human feedback (RLHF), LLMs demonstrate advanced capacity in language analysis, rationale, and interaction. While LLMs have attracted increasing amount of attention on topics such as chain-of-thought~\citep{wei_chain--thought_2022}, interpretability~\citep{zhao_explainability_2024}, and in-context learning~\citep{wei_larger_2023}, they are limited to textual data and fail to engage with a broader range of AIGC tasks.

AI-empowered agents have been devised to tackle more practical applications by equipping LLMs with task-specific modules~\citep{qian_communicative_2023}. These agents~\citep{shen_hugginggpt_2023,huang_audiogpt_2023} use the LLM to interpret a user query to some basic tasks and call task-specific modules (namely expert models) with an appropriate order. By using a modular approach, AI-driven agents are capable of solving intricate tasks without the requirement of additional training. In the audio domain, WavJourney \citep{liu_wavjourney_2023} proposed an AI-driven agent that synthesises an audio clip by connecting speech, audio, and music generative models. An audio script is created based on user instructions and compiled into an executable computer program.  The computer program then invokes various audio generative models to synthesize a recording. Despite its success, the current audio agents cannot use audio clips as input, hindering themselves from a broader range of audio generation applications. Considering the collaborative ability of LLMs and the real-world need for multimodal interactive creation, a natural question arises: \textit{can we improve audio agents with the ability of audio analysis and transformation?}

In this work, we introduce WavCraft, an AI-empowered audio agent that leverages LLMs, together with a variety of expert models, to edit and generate audio content based on human instructions and available audio materials. Given a few audio references, WavCraft analyses the content of audio references with an audio analysis module to produce audio description. The audio description and the user query are wrapped up with a pre-defined instruction template and directed to an audio programmer module. The audio programmer applies an LLM to break down a complex audio content creation task into several basic ones, and generates an executable program to invoke modules like audio expert models, DSP functions, or logic operations. Following this modular approach, WavCraft is able to assemble a variety of audio creation tools with great flexibility. The overall framework of WavCraft and its featured use cases are shown in Figure~\ref{fig:wavcraft_overview}.

Features of WavCraft include: 1) Adjustable: WavCraft can take available audio clips as raw materials and create audio content based on both user instruction and input audio. Compared with existing audio agents~\citep{liu_wavjourney_2023}, WavCraft is capable of a broader range of audio content creation, such as sound infilling and removal; 2) Modular: WavCraft can break down a comprehensive instruction into several basic audio tasks and thus handle a wide range of audio content generation tasks. In addition, the decompositional framework of WavCraft presents an explicit pathway of content creation, enhancing the explainability in the eyes of users; 3) Interactive: Exploiting the language analysis ability of the LLM, WavCraft interacts with users in multiple dialogues. During multi-round co-creation, the generated audio clips stay consistent with each other; 4) Creative: Based on the analysis of audio content and user's blueprint, WavCraft leverages the LLM to narrate a story, infers instructions for expert models, and creates the audio content that fulfill the storyline. We refer to such ability of WavCraft to generate audio content without explicit user instruction as \emph{audio scriptwriting}.

The contributions of this paper are summarised as follows:
\begin{itemize}[itemsep=0pt,parsep=8pt,leftmargin=16pt,labelsep=8pt]
    \vspace{-0.3cm}
    \item An LLM-driven audio agent named WavCraft is proposed to create or edit audio content based on user instructions and available audio clips.
    \item By coordinating various generative models, WavCraft produces audio content in a controllable manner. Our experiments demonstrate that WavCraft achieves better performance on audio generation and editing compared to current models.
    \item Additional experiments are conducted to evaluate WavCraft's ability of audio scriptwriting where models should manipulate audio content without explicit user commands. We hope this will facilitate the process of audio production.
\end{itemize}

\section{Related work} \label{sec:related_work}
\textbf{Language models for multi-modal tasks.}
Language models such as ChatGPT~\citep{openai_gpt-4_2023} and LLaMA~\citep{touvron_llama_2023} have achieved considerable progress in natural language processing. These models, featuring billions of parameters, are trained with massive high-quality training data to handle a variety of text-related tasks with a single model. To extend the open-world knowledge to more domains, subsequent works~\citep{alayrac_flamingo_2022,li_blip-2_2023} have aligned audio, video, and/or image to text and proposed multi-modal foundational models in their respective domains. While multi-modal language models achieved state-of-the-art performance in downstream tasks, most of them are restricted to common text-guided generation tasks, such as text-to-audio synthesis~\citep{liu_audioldm_2023}, where the model is not required to analyse complex user instructions.

\textbf{LLM-based agents.} 
More recently, LLM-based agents have attracted an increasing amount of attention to tackle challenging, intricate applications by integrating language models with a set of task-specific models. Toolformer~\citep{schick_toolformer_2023} was trained to decide when and where an API should be called and how to assemble outputs from different APIs. HuggingGPT~\citep{shen_hugginggpt_2023} applied ChatGPT~\citep{openai_gpt-4_2023} as a controller to allocate existing neural networks in Huggingface~\citep{shen_hugginggpt_2023}. HuggingGPT is thus capable of solving diverse AI tasks across natural language, visual, and audio domains. Meanwhile, ViperGPT~\citep{suris_vipergpt_2023}, VisProg~\citep{gupta_visual_2023}, and RVP~\citep{ge_recursive_2023} have demonstrated the promise of visual agents on image/video analysis tasks. LLaVA-Plus~\citep{liu_llava-plus_2023} extended the input query to visual domain by replacing LLM with visual language model (VLM). In the audio domain, AudioGPT~\citep{huang_audiogpt_2023} connected multiple audio neural networks and used ChatGPT to classify the user query into a predefined task. WavJourney~\citep{liu_wavjourney_2023} used an LLM to screenwrite the audio scripts and then generate audio clips by calling diverse audio generative models. Although considerable progress has been made in previous works to extend open-world knowledge in language models to multiple modalities, few of them can be prompted by non-text inputs, restricting their ability to many practical applications, especially audio editing.

\textbf{Audio Creation and Editing.} Audio creation and editing are challenging parts of generative AI since they require models to not only understand the audio content but also modify the audio conditioned on input instructions. With the development of deep learning, generative models have demonstrated remarkable capacities to synthesise speech~\citep{wang_neural_2023}, audio~\citep{liu_audioldm_2023,kreuk_audiogen_2023,borsos_audiolm_2022,team_audiobox_nodate}, and music~\citep{copet_simple_2023,agostinelli_musiclm_2023}. Existing audio generation methods are mainly dedicated to synthesising audio conditioned on different types of prompts, such as text description, voice style, and music melody. However, these methods are trained to generate audio from scratch and thus are ill-suited to editing tasks on existing audio. Recently, AUDIT~\citep{wang_audit_2023} was proposed to learn an end-to-end diffusion model to modify the audio content based on both text instructions and input audio. While AUDIT is capable of various basic editing tasks, including adding, removal, replacement, super-resolution, and infilling, it suffers from two drawbacks: 1) it does not perform well in complex editing tasks that combines these basic tasks; and 2) it cannot perform local changes on designated audio regions, limiting its application in real-world scenario. 


\begin{table*}[h]
\small
\centering
\refstepcounter{table}
\caption{List of the audio APIs and their implementation used by WavCraft.}
\label{tab:audio_modules}
\scalebox{0.8}{
\begin{tblr}{
    width=\linewidth,
    cell{2}{1}={c=5}{h},
    cell{11}{1}={c=5}{h},
    hline{1-2,11,21}={-}{},
    }
\textbf{Task}                                        & \textbf{Input}  & \textbf{Output} & \textbf{API Name} & \textbf{Model name}                                                      \\
\textit{task-specific models for audio manipulation} &             &        &             &                                                            \\
Text-to-Audio                                        & Text        & Audio  & TTA         & AudioGen \citep{kreuk_audiogen_2023} \\
Text-to-Speech~                                      & Text        & Audio  & TTS         & Bark~\footnote{https://github.com/suno-ai/bark} \\
Text-guided Source Separation                        & Text, Audio & Audio  & TSS         & AudioSep \citep{liu_separate_2023}       \\
Extract                                              & Audio       & Audio  & EXTRACT     & AudioSep \citep{liu_separate_2023}       \\
Text-to-Music                                        & Text        & Audio  & TTM         & MusicGen \citep{copet_simple_2023}       \\
Super resolution                                     & Audio       & Audio  & SR          & AudioSR~\citep{liu_audiosr_2023}                                                 \\
Drop                                                 & Text        & Audio  & DROP        & AudioSep \citep{liu_separate_2023}       \\
Inpaint                                              & Audio       & Audio  & INPANT      & AudioLDM \citep{liu_audioldm_2023}       \\
\textit{basic audio processing functions}            &             &        &             &                                                            \\
Mix                                                  & Audio       & Audio  & MIX         & \texttt{numpy.add}                                                \\
Length                                               & Audio       & Text   & LEN         & \texttt{len}                                                      \\
Concatenate                                          & Audio       & Audio  & CAT         & \texttt{numpy.concatenate}                                        \\
Clip                                                 & Audio       & Audio  & CLIP        & \texttt{numpy.ndarray}                                            \\
Adjust Volume                                        & Audio       & Audio  & ADJUST\_VOL & \texttt{torchaudio.Vol}                                           \\
Low Pass                                             & Audio       & Audio  & LOW\_PASS   & \texttt{audiomentations.LowPassFilter}                            \\
High Pass~                                           & Audio       & Audio  & HIGH\_PASS  & \texttt{audiomentations.HighPassFilter}                           \\
Room Simulate                                        & Audio       & Audio  & ROOM\_SIM   & \texttt{audiomentations.RoomSimulaor}                             \\
Impulse Response                                     & Audio       & Audio  & ADD\_RIR    & \texttt{audiomentations.ApplyImpulseResponse}                    
\end{tblr}
}
\end{table*}
\section{WavCraft} \label{sec:proposed_method}
\subsection{Overall framework} \label{subsec:overall_framework}
WavCraft is an LLM-driven system equipped with a set of task-specific audio networks, capable of audio editing and creation. The overall framework of WavCraft can be found in Figure~\ref{fig:wavcraft_overview}, highlighting three core designs: 1) \textit{audio analysis}: WavCraft initially describes the content of audio clips using natural language; 2) \textit{task decomposition}: given an user query and input audio descriptions, WavCraft formulates a set of instructions from a predefined template by prompting ChatGPT~\citep{openai_gpt-4_2023} directly; 3) \textit{code execution}: WavCraft calls the APIs of expert audio models to execute the generated computer program. We will detail these core designs in the following:

\textbf{Audio analysis.} Complex audio editing requires models to modify audio clips based on user queries and the content of input recordings. Therefore, WavCraft applies the audio analysis module to describe input audio clips in natural language. We apply an audio question and answering model to describe sounds using a template question ``write an audio caption to describe the sound''. Please note that WavCraft can integrate any models for audio question and answering~\citep{liang_acoustic_2023,deshmukh_pengi_2023}, or even an audio captioning model~\citep{mei_wavcaps_2023}, as the audio analysis module.

\textbf{Task decomposition.} The audio programmer module in WavCraft then drives the LLM to generate an executable script conditioned on both the user query and the content of input audio clips. Specifically, WavCraft fills a pre-defined template with the input query and audio description and then directs the instructions to the LLM (see more in the Appendix~\ref{appendix:prompt_template}). 
Compared with other AI programmers~\citep{gupta_visual_2023}, WavCraft generates not only the code but also the comment for each line and the audio script. We found that these comments and the audio script facilitate the audio programmer module to generate code step by step, leading to high-quality output and explainable operations. 

\textbf{Code execution.} WavCraft executes the generated scripts by calling a set of audio expert models. Table~\ref{tab:audio_modules} lists the APIs constituting WavCraft. WavCraft consists of various publicly-available expert models: AudioGen~\citep{kreuk_audiogen_2023} was used for text-to-audio generation; MusicGen~\citep{copet_simple_2023} was adopted to music synthesis due to its high-fidelity performance based on text and/or melody. For text-to-speech generation, we use Bark\footnote{https://huggingface.co/spaces/suno/bark}, a state-of-the-art model that generates matched speech conditioned on the tone, pitch, emotion, and prosody of a given voice preset. For text-guided source separation, AudioSep~\citep{liu_separate_2023} is used to separate targeted sound tracks conditioned on language queries. AudioSR~\citep{liu_audiosr_2023} and AudioLDM~\citep{liu_audioldm_2023} are used for super-resolution and audio infilling, respectively. In addition, a series of DSP modules are introduced as well. We implement the DSP modules by using torchaudio~\citep{yang_torchaudio_2022} and audiomentations~\footnote{https://github.com/iver56/audiomentations}. It is noteworthy that these task-specific modules can be easily replaced with the alternative architectures.

\subsection{Features} \label{subsec:features}
Empowered by LLMs, WavCraft is capable of intricate audio editing and creation tasks. WavCraft highlighted four advanced features:

\textbf{Modular operations.} WavCraft decomposes a user instruction into several basic tasks and thus is capable of more complex editing applications in an explainable manner.

\textbf{Controllable editing.} WavCraft translates user requests into executable lines such that it can edit the targeted attributes while keeping the rest unchanged.

\textbf{Human-AI co-creation.} WavCraft leverages large language models to edit audio in a interactive manner, facilitating human producers to create audio content through multi-round refinement. Moreover, WavCraft generates the audio script and comment lines to explain the process of audio content creation. This chain-of-thought method improves the interpretability and transparency of WavCraft.

\textbf{Audio scriptwriting.} Beyond audio content generation under explicit user guidance, WavCraft can produce the sounds in a creative approach, following a high-level outline. We name this ability to devise a plot itself and then manipulate the audio content as \textit{audio scriptwriting}. To make an audio drama, WavCraft creates a script conditioned on input audio, together with the outline, and then sonifies the script with a variety of expert models.

\section{Tasks} \label{sec:tasks}
WavCraft provides a flexible framework that can address a diverse range of audio generative and editing tasks. We evaluate WavCraft on five basic tasks, involving adding, removal, replacement, super-resolution, and infilling. We also assess the advanced features of WavCraft on complex tasks, such as controllable editing and audio scriptwriting.

\textbf{Adding.} Given two audio clips $A$ and $B$, the model is required to output a mixture $M$ by combining $A$ and $B$. Suppose $C_A$ is the caption (i.e., text description) of A, an example of the instruction can be ``Add {$C_A$} in the background of {$C_B$}".

\textbf{Removal.} Given a mixture $M$ and one of its sound tracks $A$, the model is required to output a new audio clip $B$ by removing $A$ from $M$. Suppose $C_A$ and $C_M$ are the captions of A, respectively, an example of the instruction can be ``Remove {$C_A$} from {$C_M$}".

\textbf{Replacement.} Given an audio clip $A$, a mixture $M$ and one of its sound tracks $B$, the model is required to output a new audio clip $C$ by replacing $B$ with $A$ in the same time slot. An example of the instruction can be ``Replace {$C_B$} with {$C_A$}".

\textbf{Super-resolution.} Given a low-resolution audio clip $A$, the model is required to output a new audio clip $A'$ with a higher sampling rate. An example of the instruction can be ``Increase resolution of {$A$}".

\textbf{Audio infilling.} Given an audio clip $A$ where some parts are masked, the model is required to complete the audio by filling the masked areas. An example of the instruction can be ``Inpaint {$A$}".

In addition to the basic tasks, we also evaluate the advanced features of WavCraft through a case study.

\begin{table*}[t]
\centering
\caption{Objective evaluation results on five different editing tasks.}
\label{tab:objective_eval_edit}
\begin{tabular}{@{}ccccc|cccc@{}}
\toprule
\multirow{2}{*}{Task} & \multicolumn{4}{c|}{AUDIT~\citep{wang_audit_2023}} & \multicolumn{4}{c}{WavCraft} \\
 & FAD~$\downarrow$ & IS~$\uparrow$ & KL~$\downarrow$ & LSD~$\downarrow$ & FAD~$\downarrow$ & IS~$\uparrow$ & KL~$\downarrow$ & LSD~$\downarrow$ \\ \midrule
Add                   & 9.27  & 3.87 & 3.00 & 1.95 & 0.63  & 6.05  & 1.45  & 1.59 \\
Removal              & 17.57 & 3.27 & 4.40 & 3.46 & 3.48  & 6.38  & 1.72  & 2.07 \\
Replacement           & 10.24 & 2.86 & 3.10 & 2.55 & 0.72  & 6.09  & 2.16  & 1.77 \\
Infilling             & 12.61 & 3.88 & 2.86 & 3.40 & 3.31  & 6.37  & 1.00  & 2.10 \\
Super-resolution      & 13.68 & 2.62 & 4.25 & 2.50 & 5.98  & 5.96  & 1.26  & 1.93 \\ \bottomrule
\end{tabular}%
\end{table*}

\begin{table}[t]
\centering
\caption{Objective evaluation results on the AudioCaps evaluation set.}
\label{tab:objective_eval_gen}
\begin{tabular}{@{}cccc@{}}
\toprule
\multicolumn{1}{c}{Model} & FAD~$\downarrow$ & KL~$\downarrow$ & IS~$\uparrow$ \\ \midrule
AudioLDM~\citep{liu_audioldm_2023} & 4.65 & 1.89 & 7.91 \\
WavJourney\citep{liu_wavjourney_2023} & 3.38 & \textbf{1.53} & 7.94 \\
WavCraft                  & \textbf{2.95} & 1.68 & \textbf{8.07} \\ \bottomrule
\end{tabular}
\end{table}

\section{Experiments} \label{sec:experiments}
\subsection{Experiments setup} \label{subsec:experiments_setup}
To build up WavCraft, we used the GPT-4 model~\citep{openai_gpt-4_2023} as audio programming module and LTU~\citep{gong_listen_2023} for audio analysis. We applied the publicly available models as audio expert models (shown in Table~\ref{tab:audio_modules}). We use 16 kHz sampling rate throughout the pipeline of audio editing and generation in line with the sampling rate of many integrated generative models~\citep{kreuk_audiogen_2023,copet_simple_2023}. For the volume control of the generated audio content, we adopt the Loudness Unit Full Scale (LUFS) standard~\citep{international_telecommunication_union_itu-r_2020}.

We evaluated WavCraft on audio editing and generation tasks separately. We compared WavCraft with AUDIT~\citep{wang_audit_2023}, an state-of-the-art audio editing model, on diverse downstream tasks. For text-to-audio generation, we used AudioLDM~\citep{liu_audioldm_2023} and WavJourney~\citep{liu_wavjourney_2023} for comparison. While WavJourney is also an LLM-based agent for audio content generation, it cannot take waveforms as inputs. We evaluate editing and text-to-audio generation task on AudioCaps~\citep{kim_audiocaps_2019} datasets. For editing tasks, we synthesis the database in according to~\ref{sec:tasks} while directly using the evaluation split for the text-to-audio generation task. Please note that we will refer to the synthesised audio samples as the ground truth in the following sessions.

\subsection{Evaluation metrics} \label{subsec:evaluation_metrics}
For objective evaluation, we follows the evaluation protocols of existing audio generative models~\citep{liu_audioldm_2023,liu_wavjourney_2023,wang_audit_2023} to calculate several measurements: Frechet Audio Distance (FAD), Kullback-Leibler
Divergence (KL), Inception Score (IS) and Log Spectral Distance (LSD) for evaluation. FAD measures the Frechet distance between reference and generated audio distributions of the embeddings extracted by a pre-trained VGGish model~\citep{gemmeke_audio_2017}. KL computes the similarity between logit distributions of two audio groups by using an audio tagging model, namely Patch-out Transformer~\citep{koutini_efficient_2022}. IS reflects the variety and diversity of the generated audio group. Log Spectral Distance (LSD) calculates the distance between frequency spectrograms of output samples and target samples. A lower score of FAD, KL, or LSD indicates a better audio fidelity while a higher IS indicates a more diverse audio group (and thus more desirable for generated audio). Subjective evaluation were carried out by Amazon Mechanical Turk~\footnote{https://requester.mturk.com}. We offered raters with detailed instructions and illustrative examples to ensure a thorough evaluation process. Each audio sample was rated from one to five by a minimum of 15 different raters. We collected the feedback from rates and calculate mean opinion score (MOS) to reflect the overall quality of audio samples. Likewise, we assess the ability of audio scriptwriting, rated from one to fice, from five different aspects: audio-text relevance, audio coherence, naturalness, engagement, and creativity. To extend AUDIT to the audio scriptwriting task, we use chatgpt to convert the instruction to several basic tasks and call AUDIT to execute tasks recursively. We refer to the extended AUDIT AUDIT$+$ in this paper. 

\subsection{Objective evaluation} \label{subsec:objective_evaluation}
Here we evaluate the performance of WavCraft on audio editing and text-to-audio generation tasks separately.

\textbf{Evaluation on audio editing tasks.}
Table~\ref{tab:objective_eval_edit} shows the objective results of AUDIT and the proposed WavCraft. The WavCraft achieves a better performance than AUDIT in all objective evaluation across different tasks. Compared with AUDIT, WavCraft is 8.97, 14.09, 9.52 lower in FAD on the add, removal, and replacement tasks, respectively. On the audio infilling task, WavCraft achieves the FAD score of 3.31 while having the LSD score of 1.93. 

\textbf{Evaluation on text-to-audio generation.} Table~\ref{tab:objective_eval_gen} shows the objective results of our WavCraft and the compared methods. WavCraft achieves the best FAD and IS score among the three evaluted models on the AudioCaps evaluation dataset. WavCraft also yields the KL score of 1.68, close to the performance of WavJourney~\citep{liu_wavjourney_2023}. 

\subsection{Subjective evaluation} \label{subsec:subjective_evaluation}
\begin{wrapfigure}{r}{0.5\linewidth}
    \centering
    \includegraphics[width=0.85\linewidth]{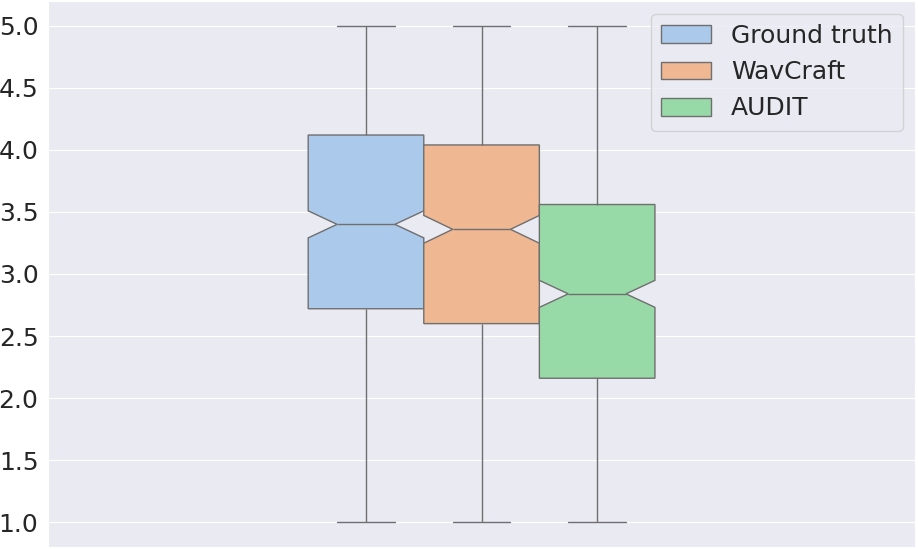}
    \caption{Overall subjective evaluation on audio editing quality by comparing the performance of the ground truth, AUDIT, and the proposed WavCraft.}
    \label{fig:overall_controllable_edit}
\end{wrapfigure}
Figure~\ref{fig:overall_controllable_edit} shows the overall subjective evaluation results by comparing the performance of the ground truth, AUDIT, and our WavCraft. MOS value of WavCraft is very close to that of the ground truth while outperform the AUDIT's by a large margin. 

\begin{figure}[htbp]
  \centering
  \begin{minipage}{0.45\linewidth}
    \centering
    \includegraphics[width=0.85\linewidth]{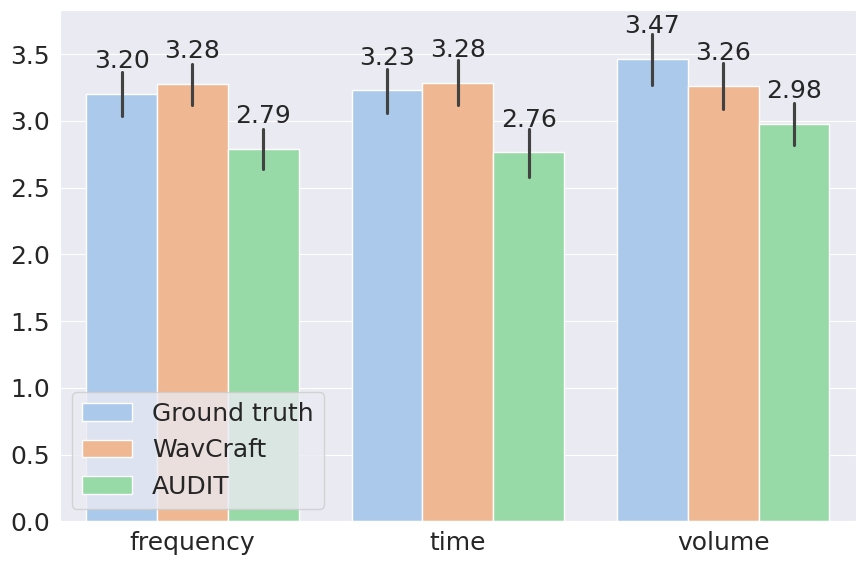}
    \caption{Subjective evaluation on the quality of edited audio obtained from the ground truth, AUDIT, and the proposed WavCraft in terms of frequency, temporal, and volume scale.}
    \label{fig:breakdown_controllable_edit}
  \end{minipage}\hfill
  \begin{minipage}{0.42\linewidth}
    \centering
    \vspace{-0.2cm}
    \includegraphics[width=0.95\linewidth]{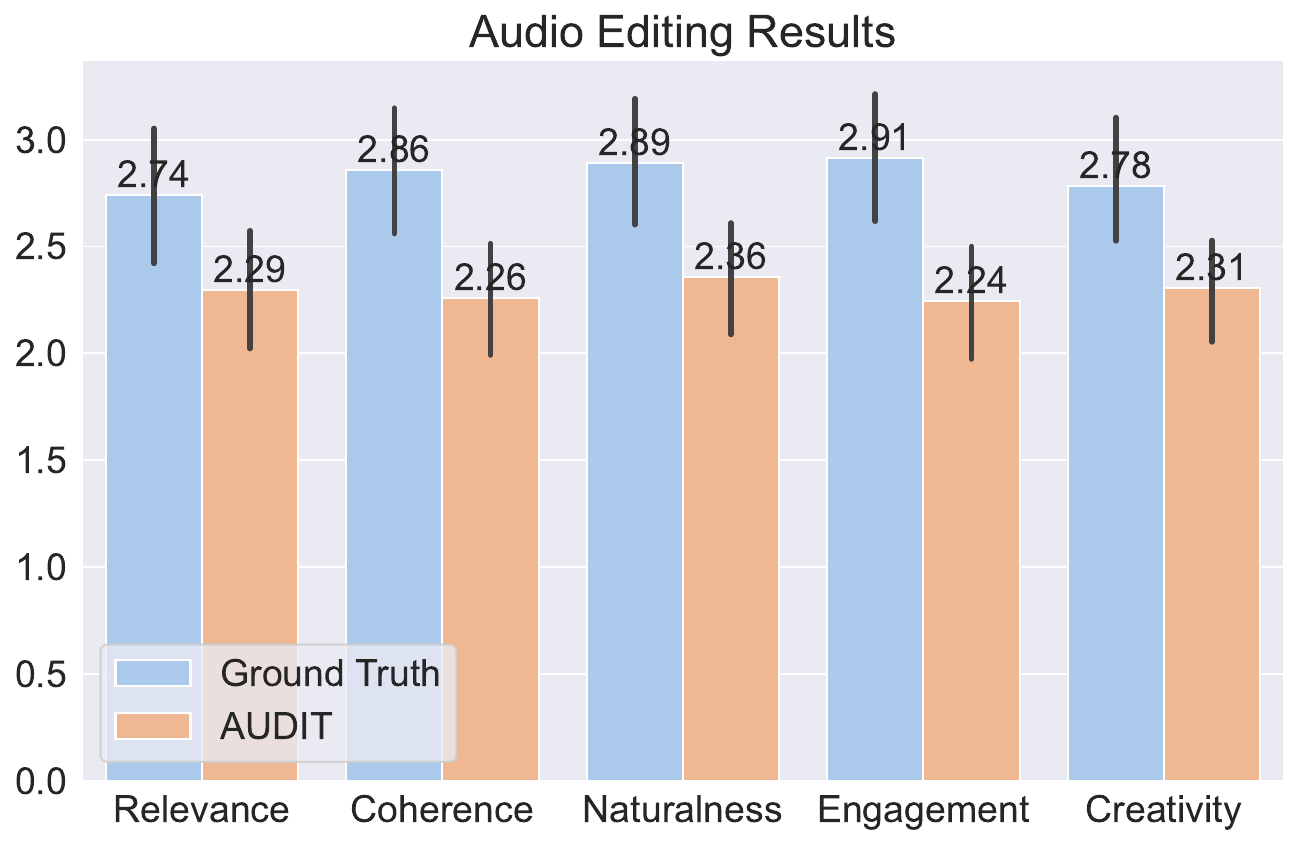}
    \caption{Comparing the ability of audio storytelling between AUDIT$+$ and the proposed WavCraft in terms of audio-text relevance, audio coherence, naturalness, engagement, and creativity.}
    \label{fig:audio_stroytelling}
  \end{minipage}
\end{figure}

Figure~\ref{fig:breakdown_controllable_edit} compares the performance of the ground truth, AUDIT, and the proposed WavCraft in terms of frequency, time, and volume. WavCraft achieves the best results on frequency and time control compared to the AUDIT and even the ground truth. We hypothesise this is partly because the raw audio material used by the ground truth is less perceptually significant than the audio segments generated by WavCraft. In addition, WavCraft achieved a better MOS value compared to AUDIT in terms of volume. 

Figure~\ref{fig:audio_stroytelling} shows the performance of WavCraft and AUDIT$+$ in terms of audio-text relevance, audio coherence, naturalness, engagement, and creativity. The WavCraft yielded a better scores than AUDIT$+$ from all aspects.

\subsection{Case study} \label{subsec:case_study}
\begin{figure*}[h]
    \centering
    \includegraphics[width=1.0\linewidth]{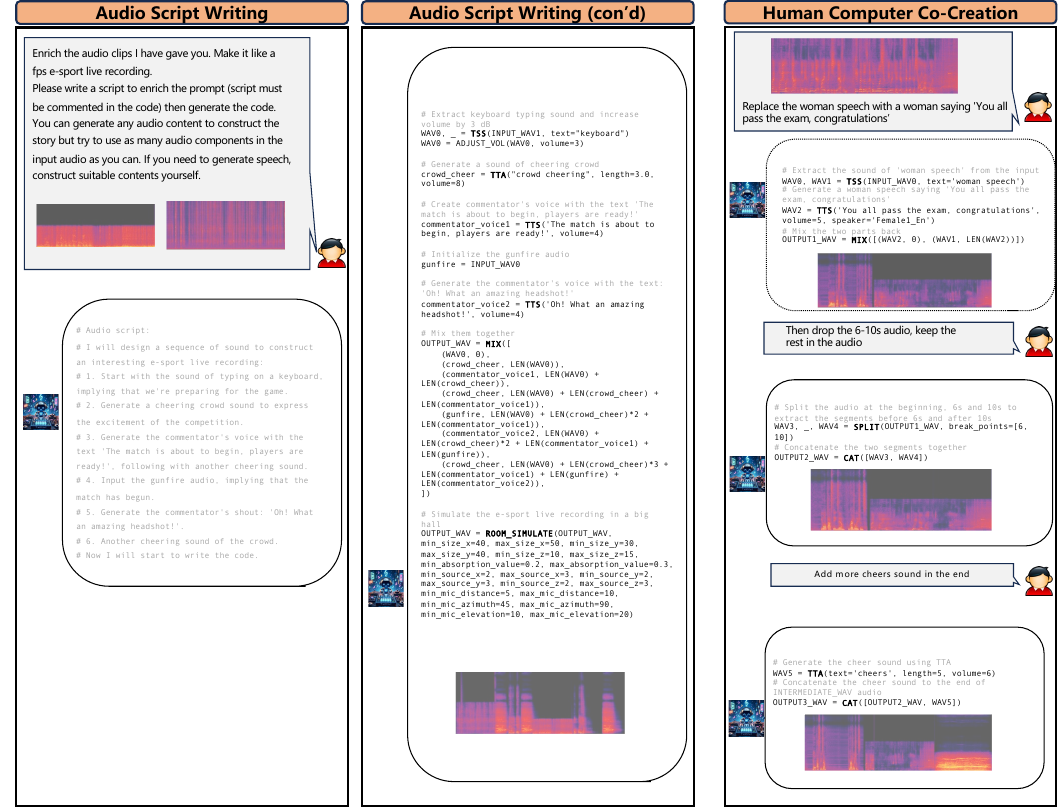}
    \caption{Case studies on audio scriptwriting and human-AI co-creation.}
    \label{fig:wavcraft_casestudy}
\end{figure*}

Leveraging LLM's ability of natural language processing, WavCraft decomposes user's requests into basic applications and thus is capable of the tasks beyond the common tasks described in Sect.~\ref{subsec:objective_evaluation}. As shown in Figure~\ref{fig:wavcraft_casestudy}, we hereby discuss the advanced features of the proposed WavCraft by studying two cases:

\textbf{Case study 1: Audio scriptwriting.} WavCraft takes two raw audio materials (i.e., the beginning and end of an audio titled "a fan heading to a soccer match field") and a user instruction as inputs. It first analyses the content of input audio and write an audio script conditioned on both the user instruction and audio descriptions. The script is written in the format of python coding and applied to allocate diverse modules, such as target source separation, text-to-audio generation models, and the room simulator, for audio editing. The output of activated modules is mixed together in line with the generated audio script. To the best of our knowledge, WavCraft is the only audio agent capable of such complex editing task without an explicit user command.

\textbf{Case study 2: Human-AI co-creation.} We illustrate how WavCraft interacts with an user during the process of audio production. WavCraft starts with a basic replacement editing task: replacing the female speech in the audio with another female speaking. After the user went through the generated recording, namely \texttt{OUTPUT1\_WAV}, the user further instructed the system to remove the audio content between 6-10s. WavCraft keeps track of the current conversation and generate \texttt{OUTPUT2\_WAV} based on the new user instruction and the previous executation lines. Likewise, WavCraft continued to generate \texttt{OUTPUT3\_WAV} in response to user's request about ``Add more cheers sound in the end'' while taking into consideration the previous instructions. Throughout the process of audio production, in addition to the generated audio, WavCraft also provide users with the executable code together with their comments. We hope this could improve the explainability of WavCraft operations. 

\section{Limitations} \label{sec:Limitations}
Despite WavCraft demonstrating a desirable ability of audio editing and generation, there still exists some limitations: 1) \textit{Audio analysis}: While WavCraft involves an audio analysis module to describe the raw materials, the performance of existing audio analysis models is limited, hindering WavCraft from precisely describing input audio with temporal relationship. 2) \textit{Inference cost}: WavCraft needs to call diverse APIs to solve a complex task, which introduces time costs during inference. Reducing the inference time will facilitate a more seamless human-AI co-creation, meeting the requirement of more practical applications.

\section{Conclusion} \label{sec:conclusion}
This work presented WavCraft, an agent system that integrates the LLM with diverse audio expert models, to create audio content conditioned on input audio clips and user queries. WavCraft decomposes an intricate editing work into individual basic audio tasks, after comprehending users' queries and the content of given recordings. The output of basic tasks is then assembled under the instructions formulated by audio programming module, contributing to the final output. Case studies conducted on several real-world scenarios have demonstrated the potential of WavCraft in audio production applications. WavCraft shows the feasibility of AIGC in the audio domain in a transparent, interpretable, and interactive manner. 

\section*{Acknowledgements} \label{sec:acknowledgements}
This work is supported by the Engineering and Physical Sciences Research Council [grant number EP/T518086/1]. E. Benetos is supported by a RAEng/Leverhulme Trust Research Fellowship [grant number LTRF2223-19-106]. H. Zhang is supported by the UKRI Centre for Doctoral Training in Artificial Intelligence and Music, funded by UK Research and Innovation [grant number EP/S022694/1]. This work is also partly supported by Engineering and Physical Sciences Research Council (EPSRC) Grant EP/T019751/1 “AI for Sound (AI4S)”, and a PhD scholarship from the Centre for Vision, Speech and Signal Processing (CVSSP) at the University of Surrey and BBC R\&D. For the purpose of open access, the authors have applied a Creative Commons Attribution (CC BY) licence to any Author Accepted Manuscript version arising.

\bibliography{refs}
\bibliographystyle{iclr2024_conference}

\newpage
\appendix
\section{Prompt template.} \label{appendix:prompt_template}
Table~\ref{tab:input_instruction1} and Table~\ref{tab:follow-up_instruction} list the instruction prepended to the user query for the first and following round dialogs, respectively. We exploits the LLM's ability of in-context learning to allocate diverse audio modules by mimicing previous examples. In the follow-up instruction, we guide LLM to pay attention to the current instruction together with the those of previous dialogs. Therefore, WavCraft can perform audio manipulation by following current request while keeping track of the previous instructions.

\begin{table*}[]
\centering
\caption{WavCraft's predifined instruction prepended to the user query.}
\label{tab:input_instruction1}
\begin{tabular}{@{}l@{}}
\toprule
\multicolumn{1}{c}{INPUT INSTRUCTION} \\ \midrule
\begin{tabular}[c]{@{}l@{}}You are an professional audio editor. Try to follow the instruction I give using several predefined tools:\\ LEN(wav) \# returns the duration of `wav` in seconds\\ MIX(wavs: list[tuple])  \# returns the mixture of the input `wavs`\\ CAT(wavs: list)  \# returns the concatenated wav using input `wavs`\\ SPLIT(wav, break\_points=list[float]) \# returns the split wavs using `break\_points`\\ ADJUST\_VOL(wav, volume: int)  \# returns the adjusted wav by `volume`\\ TTA(text: str, length: float, volume: int)  \# returns a generated audio conditioned on `text`\\ TTM(text: str, melody, length: float, volume: int)  \# returns a generated music conditioned on `text` \\ and (optional) `melody`\\ TTS(text: str, volume: int)  \# returns a generated speech conditioned on `text` and `speaker`. `speaker` \\ should be in ['Male1\_En', 'Male2\_En', 'Female1\_En', 'Female2\_En']\\ SR(wav, ddim\_steps: int, guidance\_scale: float, seed: int)  \# Returns a wav upsampled to 48kHz\\ TSS(wav, text: str)  \# returns foreground and background wav conditioned on `text`\\ ADD\_NOISE(wav, min\_snr\_db: float, max\_snr\_db: float)  \\ \# returns a generated audio mixed with gaussian noise\\ LOW\_PASS(wav, min\_cutoff\_freq: float, max\_cutoff\_freq: float, min\_rolloff: int, max\_rolloff: int)  \\ \# returns a generated audio processed by low pass filter\\ HIGH\_PASS(wav, min\_cutoff\_freq: float, max\_cutoff\_freq: float, min\_rolloff: int, max\_rolloff: int)  \\ \# returns a generated audio processed by high pass filter\\ ADD\_RIR(wav, ir)  \# returns a generated audio mixed with a given room impulse response\\ ROOM\_SIMULATE(wav, min\_size\_x: float, max\_size\_x: float, min\_size\_y: float, max\_size\_y: float, \\ min\_size\_z: float, max\_size\_z: float, min\_absorption\_value: float, max\_absorption\_value: float, \\ min\_source\_x: float, max\_source\_x: float, min\_source\_y: float, max\_source\_y: float, min\_source\_z: float, \\ max\_source\_z: float, min\_mic\_distance: float, max\_mic\_distance: float, min\_mic\_azimuth: float, \\ max\_mic\_azimuth: float, min\_mic\_elevation: float, max\_mic\_elevation: float)  \# returns a synthesized \\ audio by mixing the input `wav` with a room-specific synthesized impulse response \\ INPAINT(wav, text: str, onset: float, offset: float, duration: float)  \\ \# returns a fixed audio where the part between `onset` and `offset` has been inpainted\\ \\ \\ I will give you several examples:\\ Instruction:\\ Increase the volume of child speech by 5 dB, decrease the volume of drum by 3 dB, drop the sound of \\ machine sound.\\ Code:\\ \# Separate the sound of 'child speech' from the mixture and return both 'child speech' and the \\ background sounds\\ WAV0, WAV1 = TSS(INPUT\_WAV0, text="child speech")\\ \# Separate the sound of 'drum' from the mixture and return both 'drum' and the background sounds\\ WAV2, WAV3 = TSS(WAV1, text="drum")\\ \# Drop the sound of 'machine sound' from the mixture\\ \_, WAV3 = TSS(WAV3, text="machine sound")\\ \# Increace the volume of "child speech" by 5dB\\ WAV0 = ADJUST\_VOL(WAV0, volume=5)\\ \# Decrease the volume of 'drum' by 5dB\\ WAV2 = ADJUST\_VOL(WAV2, volume=-3)\\ \# Mix the resulted sounds together\\ OUTPUT\_WAV = MIX([(WAV0, 0), (WAV2, 0), (WAV3, 0)])\\ \end{tabular} \\ \bottomrule
\end{tabular}%
\end{table*}
\begin{table*}[]
\centering
\label{tab:input_instruction2}
\begin{tabular}{@{}l@{}}
\toprule
\multicolumn{1}{c}{INPUT INSTRUCTION (con'd)} \\ \midrule
\begin{tabular}[c]{@{}l@{}}Instruction:\\ Extract 1-5s of the first audio with a low-pass filter to simulate the sound coming from inside a building. \\ Replace male speech with dog barking in the second audio. Upsample the mix. \\ Code:\\ \\ \# Truncate the sound between 1s and 5 s\\ \_, WAV0, \_ = SPLIT(INPUT\_WAV0, break\_points=[1, 5])\\ \# Add a low-pass filter\\ WAV0 = LOW\_PASS(WAV0, min\_cutoff\_freq=300.0, max\_cutoff\_freq=800.0, \\ min\_rolloff=6, max\_rolloff=12)\\ \# Extract the sound of 'male speech' from the truncated sound\\ WAV1, WAV2 = TSS(INPUT\_WAV1, text="male speech")\\ \# Generate the sound of 'dog barking' with the same length with the sound of 'male speech'\\ WAV3 = TTA(text="dog barking", length=LEN(WAV1), volume=4)\\ \# Combine the sounds by mixing them together\\ MIXTURE\_WAV = MIX([(WAV3, 0), (WAV2, 0), (WAV0, 0)])\\ \# Perform super-resolution on the mixture of sounds\\ OUTPUT\_WAV = SR(MIXTURE\_WAV)\\ \\ Instruction:\\ Isolate train sound in the input audio, apply a high-pass filter and increase the volume by 3 dB. \\ Repeat it five times to simulate a longer train passing.\\ Code:\\ \# Extract the sound of a train from the audio\\ WAV0, \_ = TSS(INPUT\_WAV0, text="train")\\ \# Apply a high-pass filter to reduce low-frequency noise\\ FILTERED\_WAV0 = HIGH\_PASS(WAV0, min\_cutoff\_freq=500.0, max\_cutoff\_freq=1000.0, \\ min\_rolloff=6, max\_rolloff=12)\\ \# Increase the volume by 3 dB\\ FILTERED\_WAV0 = ADJUST\_VOL(FILTERED\_WAV0, volume=3)\\ \# Concatenate the filtered train sound three times\\ OUTPUT\_WAV = CAT([FILTERED\_WAV0] * 5)\\ \\ Instruction:\\ Extract the hammer sound from the first audio, and truncate it from the start towards 2 second. \\ Remove the sound of baby crying in the second audio, and then decrease the volume by 1 dB. \\ Mix two audio together, and the second sound should begin from 1 second. Add a reverb effect \\ to the mixture sound using the third audio.\\ Code:\\ \# Extract the hammer sound from the first audio\\ WAV0, \_ = TSS(INPUT\_WAV0, text="hammer")\\ \# Truncate from the start towards 2 second\\ WAV0, \_ = SPLIT(WAV0, break\_points=[2])\\ \# Drop the sound of baby crying in the second audio\\ \_, WAV1 = TSS(INPUT\_WAV1, text="baby crying")\\ \# Decrease the volume by 1 dB\\ WAV1 = ADJUST\_VOL(WAV1, volume=-1)\\ \# Mix the ouput sounds together\\ MIXED\_WAV = MIX([(WAV0, 0), (WAV1, 1)])\\ \# Add a reverb effect using room impulse response\\ OUTPUT\_WAV = ADD\_RIR(MIXED\_WAV, ir=INPUT\_WAV2)\end{tabular} \\ \bottomrule
\end{tabular}%
\end{table*}

\begin{table*}[]
\centering
\label{tab:input_instruction3}
\begin{tabular}{@{}l@{}}
\toprule
\multicolumn{1}{c}{INPUT INSTRUCTION (con'd)} \\ \midrule
\begin{tabular}[c]{@{}l@{}}Instruction:\\ Inpaint the first audio between 2s and 5s with the text "a car passing by with rain falling". \\ Generate a 10s long jazz music piece with the second audio as melody, then mix it with the \\ sound of rain from the first, starting at 3s into the jazz music. \\ Code:\\ \# Inpaint the first audio between 2s and 5s with the text "a car passing by with rain falling"\\ WAV0 = INPAINT(INPUT\_WAV0, text="a car passing by with rain falling", onset=2, \\ offset=5, duration=LEN(INPUT\_WAV0))\\ \# Generate a 10-second jazz music piece\\ WAV1 = TTM(text="jazz", melody=INPUT\_WAV1, length=10.0, volume=5)\\ \# Extract the sound of rain from the audio file\\ WAV0, \_ = TSS(WAV0, text="rain")\\ \# Mix the jazz music with the rain sound, starting the rain at 3 seconds\\ OUTPUT\_WAV = MIX([(WAV0, 0), (WAV1, 3)])\\ \\ Instruction:\\ Remove wind sound from an outdoor recording. Generate a 5-second saxophone music \\ with happy mood followed by "Bravo". Mix the generated sound with the outdoor \\ recording and simulate the mixture in a small room with high absorption.\\ Code:\\ \# Drop the sound of wind from the original recording\\ \_, WAV0 = TSS(INPUT\_WAV0, text="wind")\\ \# Generate a 5-second saxophone music with happy mood followed by a male speech \\ "Bravo".\\ WAV1 = TTM(text="happy saxophone", length=5.0, volume=4)\\ \# Generate a speech "Bravo"\\ WAV2 = TTS("Bravo", volume=5)\\ \# Concatenate the generated sound together\\ CONCAT\_WAV = CAT([WAV1, WAV2]) \\ \# Mix the generated sound with the background sound\\ MIXED\_WAV = MIX((WAV0, 0), (CONCAT\_WAV, 0))\\ \# Simulate the recording in a small room with high absorption\\ OUTPUT\_WAV = ROOM\_SIMULATE(MIXED\_WAV, min\_size\_x=3, max\_size\_x=4, \\ min\_size\_y=3, max\_size\_y=4, min\_size\_z=2.5, max\_size\_z=3, min\_absorption\_value=0.7, \\ max\_absorption\_value=0.9, min\_source\_x=1, max\_source\_x=1.5, min\_source\_y=1, \\ max\_source\_y=1.5, min\_source\_z=1, max\_source\_z=1.5, min\_mic\_distance=1, \\ max\_mic\_distance=1.5, min\_mic\_azimuth=45, max\_mic\_azimuth=90, \\ min\_mic\_elevation=20, max\_mic\_elevation=30)\end{tabular} \\ \bottomrule
\end{tabular}%
\end{table*}

\begin{table*}[]
\centering
\caption{WavCraft's follow-up instruction to make sure the consistency of generated audio within the multi-round dialog between human and AI.}
\label{tab:follow-up_instruction}
\begin{tabular}{@{}l@{}}
\toprule
\multicolumn{1}{c}{FOLLOW-UP INSTRUCTION} \\ \midrule
\begin{tabular}[c]{@{}l@{}}Regenerate the code by appending the new instruction to the previous instructions. \\ The code must start with the provided audio (e.g., INPUT\_WAV0) and cannot take \\ the output from previous phase (i.e., `OUTPUT\_WAV`) as a known input. The new \\ instruction is:\end{tabular} \\ \bottomrule
\end{tabular}%
\end{table*}

\end{document}